\documentclass[twocolumn,amsmath,amssymb,aps,prd,nofootinbib]{revtex4-2}

\usepackage{graphicx}
\usepackage{color}
\usepackage{multirow}
\usepackage[normalem]{ulem}

\begin{document}

\title{Coupled quintessence inspired by warm inflation}

\author{Paulo M.~S\'a}

\email{pmsa@ualg.pt}

\affiliation{
Departamento de F\'\i sica, Faculdade de Ci\^encias e Tecnologia, 
Universidade do Algarve, 8005-139 Faro, Portugal\\}

\affiliation{Instituto de Astrof\'\i sica e Ci\^encias do Espa\c co,
Faculdade de Ci\^encias da Universidade de Lisboa, Edif\'\i cio C8,
Campo Grande, 1749-016 Lisbon, Portugal}

\begin{abstract}
We investigate a coupled quintessence cosmological model in which a dark-energy
scalar field with an exponential potential interacts directly with a dark-matter
fluid through a dissipative term inspired by warm inflation. The evolution
equations of this model give rise to a three-dimensional dynamical system for
which a thorough qualitative analysis is performed for all values of the relevant
parameters. We find that the model is able to replicate the observed sequence of
late-time cosmological eras, namely, a long enough matter-dominated era followed
by a present era of accelerated expansion. In situations where there is a
significant transfer of energy from dark energy to dark matter, temporary
scaling-type solutions may arise, but, asymptotically, all solutions are
dominated by dark energy. 
\end{abstract}

\maketitle

\section{Introduction}

One of the most remarkable developments in modern cosmology was the discovery,
in the late 1990s, that the Universe is undergoing a period of accelerated
expansion \cite{riess-1998,perlmutter-1999} due to an unknown form of energy
called dark energy.
Today, this acceleration is firmly established by several other independent
precise cosmological observations \cite{astier-2012}.
In particular, recent measurements of baryon acoustic oscillations from the
first year of observations from the Dark Energy Spectroscopic Instrument
(DESI) confirm, with unprecedented precision, the accelerated expansion
of the Universe \cite{desi-2024}.

The simplest candidate for dark energy is a cosmological constant $\Lambda$,
accounting for about $68\%$ of the current total energy density of the
Universe \cite{planck-2018}.
The other components of the standard cosmological model---dark matter,
baryonic matter, and radiation---account for about $27\%$, $5\%$, and 
$0.005\%$ of the total energy density, respectively
 \cite{fixsen-2009,planck-2018}.

A theoretically appealing alternative is to consider that the role of dark
energy is played not by a cosmological constant but rather by a dynamic
scalar field whose potential energy dominates the current phase of the
evolution of the Universe \cite{caldwell-1998},
inducing the observed cosmic acceleration.
Such a possibility seems to be preferred --- albeit moderately
---  by DESI’s first-year results when they are combined with data from other studies \cite{desi-2024}.
This quintessence scalar field can in principle be directly coupled to dark
matter---whose precise nature is also unknown \cite{bertone-2018}---giving
rise to coupled quintessence \cite{amendola-2000}.

If the nature of dark energy and dark matter is currently unknown, much more
so is the form of an eventual non-gravitational coupling between them.
This circumstance has allowed for great freedom in the construction of
coupled quintessence models (also known as interacting dark energy models)
\cite{amendola-1999,holden-2000,billyard-2000,amendola-2000,zimdahl-2001,
tocchini-valentini-2002,gumjudpai-2005,zhang-2005,capozziello-2006,
boehmer-2008,henriques-2009,lopez-honorez-2010,bertolami-2012,amendola-2014, 
tzanni-2014,bruck-2016,nunes-2016,barros-2019,odintsov-2019,pan-2019,
sa-2020a,sa-2020b,oikonomou-2021,sa-2021,waeming-2022,fonseca-2022,
potting-2022,agostino-2022,bruck-2023,yang-2023}.

A popular choice in the literature identifies dark matter with a
pressureless non-relativistic perfect fluid and considers the potential of
the dark-energy scalar field $\phi$ to be of the exponential type
and the interaction term between the dark components to be of the form 
$Q\propto\rho_{\texttt{DM}} \dot{\phi}$, where $\rho_{\texttt{DM}}$ is the
energy density of the dark-matter fluid and an overdot denotes a derivative
with respect to cosmic time $t$.
Such an interaction term, which has been motivated by scalar-tensor theories,
allows for the existence of late-time accelerated scaling solutions
\cite{amendola-1999,holden-2000,billyard-2000}.
Because of its potential to address the cosmological coincidence 
problem \cite{steinhardt-1997} and the Hubble constant tension 
\cite{valentino-2021,riess-2024},
this coupled quintessence model has attracted substantial attention over
the years.

More recently, a generalized interaction term of the form
$Q\propto \rho_{\texttt{DM}} \dot{\phi} C(\phi)$, where 
$C(\phi) \propto \phi^n$ and $n$ is a positive integer,
was considered \cite{potting-2022}.
With such a coupling between dark energy and dark matter, the quintessence
model no longer admits scaling attractor solutions, but, interestingly,
for certain values of a relevant parameter, during the approach to the 
dark-energy-dominated final state, the solution mimics an accelerated
scaling solution.
Naturally, this result raises the question of how common accelerated scaling
solutions are in coupled quintessence models.
We deem this issue to deserve further investigation.

In this article, we investigate a coupled quintessence model with an
interaction term $Q$ inspired by warm inflation.

According to the warm-inflation paradigm \cite{berera-1995}, energy is
continuously transferred from the inflaton field $\psi$ to a radiation bath,
thus ensuring that the energy density of the latter, $\rho_\texttt{R}$,
is not diluted during the inflationary expansion and that a smooth transition
to a radiation-dominated era can occur without the need for a separate
postinflationary reheating phase (for recent reviews on warm inflation,
see Refs.~\cite{kamali-2023,berera-2023}).
As a result, during the inflationary period, the evolution equations for
the inflaton field and radiation require an extra dissipative term,
becoming
\begin{gather}
 \ddot{\psi}+ 3H\dot{\psi}+\frac{\partial V}{\partial\psi}=-\Gamma\dot{\psi},
\\
 \dot{\rho}_\texttt{R}+4H\rho_\texttt{R}=\Gamma \dot{\psi}^2,
\end{gather}
where $H$ is the Hubble parameter, $V=V(\psi)$ the potential of the inflaton
field, and $\Gamma$ the so-called dissipation coefficient, which, in general,
is a function of the inflaton field and the temperature of the radiation bath,
$\Gamma=\Gamma(\psi,T)$.

The above described warm-inflation paradigm can be realized in
realistic cosmological models, yielding results for the scalar spectral
index  $n_s$ and the tensor-to-scalar ratio $r$ consistent with Planck
observations (see \cite{kamali-2023,berera-2023} and references therein).
Furthermore, it has been shown that warm inflation is favored by the
de Sitter swampland conjectures 
\cite{motaharfar-2019,das-2019a,das-2019b,bertolami-2022}.

Warm inflation-type dissipation processes can be present at later stages of
the Universe's evolution, giving rise, in particular, to a direct coupling
between dark energy and dark matter. Such a possibility was recently implemented
in the context of a steep-potential quintessence inflationary model,
allowing for the unification of early and late stages of the evolution of
the Universe through dissipative effects \cite{lima-2019,das-2023}.
Here, we consider a coupled quintessence model in which,
at late times, a dark-energy scalar field $\phi$ interacts directly with a
dark-matter fluid through a dissipative term of the type
$Q=\Gamma \dot{\phi}^2 $.
As a first approach, we choose the dissipation coefficient $\Gamma$
to be constant, leaving more complex cases for future publications.

This article is organized as follows.
In Sec.~\ref{Sec-2}, we present our coupled quintessence cosmological
model and write the corresponding evolution equations as a three-dimensional
dynamical system.
In Sec.~\ref{Sec-3}, we investigate the stability properties of this
dynamical system, identify the global attractors, and describe the
trajectories that correspond to the relevant late-time cosmological solutions.
Finally, in Sec.~\ref{Sec-conclusions}, we present our conclusions.

\section{The coupled quintessence cosmological model\label{Sec-2}}

Assuming a flat Friedmann--Lema\^{\i}tre--Robertson--Walker metric,
\begin{equation}
ds^2=-dt^2+a^2(t)d\Sigma^2,
\end{equation}
where $a(t)$ is the scale factor and $d\Sigma^2$ is the metric of the
three-dimensional Euclidean space, the evolution equations of the coupled
quintessence cosmological model are given by
\begin{gather}
H^2 = \frac{\kappa^2}{3} \left( \frac{\dot{\phi}^2}{2} + V 
+\rho_\texttt{DM} \right),   \label{Friedmann 1}
\\
\dot{H} = -\frac{\kappa^2}{2}  \left(\dot{\phi}^2 + \rho_\texttt{DM} \right),
\label{dotH}
\\
\ddot{\phi} + 3H\dot{\phi} +\frac{\partial V}{\partial\phi}
= \frac{Q}{\dot{\phi}},   \label{ddotphi 1}
\\
\dot{\rho}_\texttt{DM}+3H\rho_\texttt{DM} = -Q.   \label{dotrho}
\end{gather}

In the above equations, $\phi$ is the quintessence dark-energy scalar
field with potential $V(\phi)$, $\rho_\texttt{DM}$ is the energy density of
a pressureless dark-matter fluid, $H\equiv\dot{a}/a$ is the Hubble parameter, 
$Q$ is the interaction term between dark energy and dark matter,
overdots denote derivatives with respect to cosmic time~$t$, and we use
the notation $\kappa\equiv \sqrt{8\pi G}=\sqrt{8\pi}/m_\texttt{P}$,
where $G$ is the gravitational constant and $m_\texttt{P}$ is the Planck mass.

Defining the energy density and pressure of the scalar field $\phi$ as
\begin{equation}
 \rho_\phi=\frac{\dot{\phi}^2}{2}+V(\phi) \quad \mbox{and} \quad
 p_\phi=\frac{\dot{\phi}^2}{2}-V(\phi),
 \label{p-rho_phi}
\end{equation}
equation~(\ref{ddotphi 1}) can be written as
\begin{equation}
 \dot{\rho}_\phi+3H(\rho_\phi+p_\phi) = Q.
\end{equation}

In the evolution equations (\ref{Friedmann 1})--(\ref{dotrho}),
for the sake of simplicity, ordinary baryonic matter was neglected since 
it makes a small contribution to the total energy density of matter.
Additionally, radiation was also excluded since we are only interested in
late-time cosmological solutions.

At present, no fundamental underlying theory specifies the exact form of
the interaction term $Q$ between dark energy and dark matter. Hence, any
approach to this problem is necessarily phenomenological, and the selection
of the most suitable interaction model will be decided, ultimately,
by precise cosmological observations.

In this spirit, one could consider the energy transfer between the dark
components to be dictated by nonlocal quantities, for instance, by the
Universe's expansion rate $H$, yielding an interaction term of the form
$Q\propto H f(\rho_\phi,\rho_{\texttt{DM}})$, where $f$ is some function
of $\rho_\phi$ and $\rho_{\texttt{DM}}$. Such an approach has been rather
popular in literature (for a review, see Refs.~\cite{bolotin-2015,wang-2016}.

Alternatively, we could relate this energy transfer to local dissipative
effects as in the warm inflationary scenario \cite{kamali-2023,berera-2023}.
This is our option in this article.
More specifically, we choose the coupling between dark energy and dark
matter to be of the form
\begin{equation}
	Q=\Gamma \dot{\phi}^2,
	\label{interaction term}
\end{equation}
where $\Gamma$ is a dissipation coefficient determined only
by local properties of the dark-sector interactions.
Here, we are assuming that the dissipative processes occurring
in the early Universe are also present at later stages of evolution, namely,
during the matter-dominated era and the current era of accelerated expansion
driven by the potential energy of the quintessence field.
This is a natural assumption.
Indeed, if seemingly disparate phenomena, like inflation, dark matter,
and dark energy, can be unified under the same theoretical framework (for
such a triple unification, see, for instance, Ref.~\cite{sa-2020b}, one
could also expect the interactions between the underlying particles/fields
to somehow remain active during all stages of the Universe's evolution.
These interactions would, then, give rise to the interaction term
(10) with a dissipation coefficient $\Gamma$
depending both on the scalar field $\phi$ and the dark-matter energy
density $\rho_\texttt{DM}$ \cite{lima-2019}.

In this article, as a first approach, we choose $\Gamma$ to be a constant
(with dimension of mass), leaving more general dissipative coefficients
for future work.

Note that, with an interaction term of the form (\ref{interaction term}),
the evolution equation for the quintessence scalar field (\ref{ddotphi 1})
can be written as 
\begin{equation}
\ddot{\phi} + \left(3H-\Gamma\right) \dot{\phi}
	+ \frac{\partial V}{\partial\phi} = 0,
\end{equation}
revealing that the direct interaction between the dark components modifies
the so-called ``Hubble friction" term.
In what follows, we will consider both possibilities that this term is
enhanced or diminished due to the direct interaction with the dark-matter
fluid.

In what concerns the potential of the dark-energy scalar field, we choose it
to be of the exponential form,
\begin{equation}
V(\phi) = V_a e^{-\mu\kappa\phi},
 \label{Potential}
\end{equation}
where $V_a$ is a positive constant with dimension $(\mbox{mass})^4$
and $\mu$ is a dimensionless constant.

To study the cosmological solutions of the system of 
equations (\ref{Friedmann 1})--(\ref{dotrho}) we resort
to the powerful method of qualitative analysis of dynamical systems.

To this end, we introduce the dimensionless variables
\begin{equation}
	x=\frac{\kappa \dot{\phi}}{\sqrt6 H}, \quad
	y=\frac{\kappa\sqrt{V}}{\sqrt3 H}, \quad \mbox{and} \quad
	z=\frac{H_*}{H \Omega_\texttt{DM}+H_*},
	\label{xyz}
\end{equation}
as well as a new time variable $\tau$, defined as
\begin{equation}
	\frac{d\tau}{dt}=\frac{H}{1-z},
	\label{tau}
\end{equation}
where $\Omega_\texttt{DM} \equiv \kappa^2\rho_\texttt{DM}/(3H^2)$ is the 
dark-matter density parameter and $H_*$ is a positive constant with
dimension of mass (the dimensionless variable $z$ should not
be confused with the cosmological redshift, defined as $a_0/a(t)-1$, where 
$a_0$ is the present-time value of the scale factor).
Note that from Eq.~(\ref{Friedmann 1}) it immediately
follows
\begin{equation}
	\Omega_\texttt{DM}=1-x^2-y^2.
	\label{Omega-matter}
\end{equation}

Before proceeding, some comments are in order about the choice of the
new variables $z$ and $\tau$ (the choice of $x$ and $y$ is the usual
one \cite{copeland-1998}). 
The interaction term between dark energy and dark matter, given
by Eq.~(\ref{interaction term}), cannot be written as a function of the
variables $x$ and $y$ only; therefore, an extra variable $z$ is required to
close the dynamical system.
In choosing it, we must ensure that the surface $x^2+y^2=1$ becomes an
invariant manifold of the dynamical system, so that no trajectory
can cross this surface and enter the region of (unphysical) negative values of
$\Omega_\texttt{DM}$.
In addition, we also want to compactify the phase space in the $z$
direction, say between $z=0$, corresponding to $H=+\infty$, and $z=1$,
corresponding to $H=0$.
The simplest choice of $z$ that satisfies these requirements is given in
Eq.~(\ref{xyz}). However, with such $z$, the interaction term
appearing in the evolution equation for $x$ becomes proportional to
$(1-z)^{-1}$ and, consequently, diverges as $z\rightarrow 1$.
We could study the properties of the dynamical system on the $z=1$ plane 
by just considering the divergent term and neglecting the others.
Instead, we opt to remove the singularity by choosing $d\tau$
proportional to $(1-z)^{-1}$, which amounts to multiplying the right-hand side
of the evolution equations by $1-z$.
This procedure allows us to study the dynamical system on the $z=1$ plane
(without any divergent terms) and simultaneously preserves the stability
properties in other regions of phase space.

In the variables $x$, $y$, $z$, and $\tau$, the evolution equations
(\ref{Friedmann 1})--(\ref{dotrho}) of the coupled quintessence
cosmological model reduce to a three-dimensional dynamical system, namely,
\begin{subequations}\label{DSn}
\begin{align}
 x_\tau = &  \bigg[ -3x + \frac{\sqrt{6}}{2} \mu y^2
 + \frac32 x(1+x^2-y^2) \bigg] (1-z) \nonumber 
\\
  &+ \alpha x (1-x^2-y^2)z, \label{DSn-x}
\\
 y_\tau = & \bigg[ -\frac{\sqrt{6}}{2} \mu x
 + \frac32 (1+x^2-y^2) \bigg] y(1-z),
      \label{DSn-y}
\\
 z_\tau = & \bigg[ \frac32 (1-x^2+y^2)(1-z)
 + 2\alpha x^2 z \bigg] z(1-z)
 \label{DSn-z},
\end{align}
\end{subequations}
where $\alpha=\Gamma/H_*$ is a dimensionless constant parameterizing
the energy exchange between dark matter and dark energy.

Inspection of this dynamical system reveals that the surfaces $y=0$,
$z=0$, and $z=1$ are invariant manifolds.
Furthermore, from the evolution equation for the dark-matter density
parameter, obtained from Eqs.~(\ref{Omega-matter}),
(\ref{DSn-x}), and (\ref{DSn-y}),
\begin{equation}
	\Omega_{\texttt{DM},\tau}=\Omega_\texttt{DM} \left[ 3(x^2-y^2) (1-z)
	-2\alpha x^2 z
	 \right],
\end{equation}
we conclude that the surface $x^2+y^2=1$ ($\Omega_\texttt{DM}=0$)
is also an invariant manifold.
Taking into account that $\Omega_\texttt{DM}$ is, by definition,
non-negative and restricting ourselves to expanding cosmologies, 
the phase space of the dynamical system~(\ref{DSn}) is then the half-cylinder
$\{(x, y, z)\,|\, x^2+y^2 \leq 1, y \geq 0,0 \leq z \leq 1\}$.

In what concerns the parameter space, let us point out that the dynamical
system~(\ref{DSn}) is invariant under the transformation $x\rightarrow -x$
and $\mu\rightarrow-\mu$, implying that, without any loss of generality,
the parameter $\mu$ can be assumed to be positive.
Furthermore, we assume that the parameter $\alpha$ can be either positive
(energy is transferred from the dark-matter fluid to the dark-energy scalar
field) or negative (energy is transferred in the opposite direction). 
Altogether, this means that the parameter space of our coupled
quintessence model
is $\{(\alpha,\mu)\,|\, \alpha\neq0, \mu>0\}$.

In the dimensionless variables $x$, $y$, and $z$, the dark-energy density
parameter and the effective equation-of-state parameter are given by
\begin{equation}
 \Omega_\phi \equiv \frac{\kappa^2}{3H^2} \rho_\phi =x^2+y^2
\end{equation}
and
\begin{equation}
 w_\texttt{eff} \equiv \frac{p_\phi}{\rho_\texttt{DM}+\rho_\phi}
 =x^2-y^2,
\end{equation}
respectively, where $\rho_\phi$ and $p_\phi$ are defined in 
Eq.~(\ref{p-rho_phi}).
The latter quantity may take values in the range $[-1,1]$.
When the evolution of the Universe is completely dominated by the potential
$V(\phi)$, the effective equation-of-state parameter equals $-1$, and one
recovers the cosmological constant case.
When the dynamics is dominated by the dark-matter fluid or the scalar-field 
kinetic term one obtains $w_\texttt{eff}=0$ or $w_\texttt{eff}=1$,
respectively.
Expansion is accelerated if $-1\leq w_\texttt{eff}<-1/3$.

The coupled phantom cosmological model inspired by warm inflation,
in which $w_\texttt{eff}$ may be less than $-1$, was studied in
Ref.~\cite{halder-2024}.

\section{Cosmological solutions\label{Sec-3}}

The dynamical system~(\ref{DSn}) has two critical lines and, depending
on the values of the parameters $\alpha$ and $\mu$, up to six critical points. 
Table~\ref{Table:properties CP} summarizes their properties, namely,
the conditions for their existence and the physical behavior of the solutions
in their vicinity, while Table~\ref{Table:stability CP} summarizes the
stability properties of the critical points and lines.

\begin{table*}[t]
\begin{tabular}{cccccc}
\hline
Critical point/line
  & \quad Existence
  & \quad $\Omega_\phi$
  & \quad $\Omega_\texttt{DM}$
  & \quad $w_{\rm eff}$
  & \quad Acceleration \\ \hline \noalign{\vskip 1mm}
$A(1,0,0)$
  & \quad $\mu>0$, $\alpha\neq0$
  & \quad $1$
  & \quad $0$
  & \quad $1$
  & \quad no \\ \noalign{\vskip 1mm}
$B(-1,0,0)$
  & \quad $\mu>0$, $\alpha\neq0$
  & \quad $1$
  & \quad $0$
  & \quad $1$
  & \quad no \\ \noalign{\vskip 1mm}
$C(0,0,0)$
  & \quad $\mu>0$, $\alpha\neq0$
  & \quad $0$
  & \quad $1$
  & \quad $0$
  & \quad no \\ \noalign{\vskip 1mm}
$D\left(\frac{\mu}{\sqrt6},\sqrt{1-\frac{\mu^2}{6}},0\right)$
  & \quad $0<\mu\leq\sqrt6$, $\alpha\neq0$
  & \quad $1$
  & \quad $0$
  & \quad $-1+\frac{\mu^2}{3}$
  & \quad $0<\mu<\sqrt2$ \\ \noalign{\vskip 1mm}
$E\left(\frac{\sqrt6}{2\mu},\frac{\sqrt6}{2\mu},0\right)$
  & \quad $\mu\geq\sqrt3$, $\alpha\neq0$
  & \quad $\frac{3}{\mu^2}$
  & \quad $1-\frac{3}{\mu^2}$
  & \quad $0$
  & \quad no \\ \noalign{\vskip 1mm}
$F\left(\frac{\mu}{\sqrt6},\sqrt{1-\frac{\mu^2}{6}},
\frac{3(\mu^2-6)}{(2\alpha+3)\mu^2-18}\right)$
  & \quad $0<\mu<\sqrt{6}$, $\alpha<0$ \mbox{or} $\mu=\sqrt6$, $\alpha\neq0$
  & \quad $1$
  & \quad $0$
  & \quad $-1+\frac{\mu^2}{3}$
  & \quad $0<\mu<\sqrt2$ \\ \noalign{\vskip 1mm}
$G(0,y,1)$
  & \quad $\mu>0$, $\alpha\neq0$, $0\leq y\leq1$
  & \quad $y^2$
  & \quad $1-y^2$
  & \quad $-y^2$
  & \quad $\frac{1}{\sqrt3}<y\leq1$ \\ \noalign{\vskip 1mm}
$H\left(x,\sqrt{1-x^2},1\right)$
  & \quad $\mu>0$, $\alpha\neq0$, $-1\leq x\leq1$
  & \quad $1$
  & \quad $0$
  & \quad $-1+2x^2$
  & \quad $|x|<\frac{1}{\sqrt3}$ \\ \noalign{\vskip 1mm}
    \hline
\end{tabular}
\caption{\label{Table:properties CP} Properties of the critical points and
critical lines of the dynamical system~(\ref{DSn}).
$\Omega_\phi$, $\Omega_\texttt{DM}$, and $w_\texttt{eff}$ are, respectively,
the dark-energy density parameter, the dark-matter density parameter,
and the effective equation-of-state parameter. Accelerated expansion of the
Universe occurs for
$w_\texttt{eff}<-1/3$.}
\end{table*}

\begin{table*}[t]
\begin{tabular}{ccc}
\hline
Critical point/line
  & \quad Eigenvalues
  & \quad Stability \\ \hline \noalign{\vskip 1mm}
$A(1,0,0)$
  & \quad $\left\{3,0,3-\frac{\sqrt6}{2}\mu\right\}$
  & \quad no \\ \noalign{\vskip 1mm}
$B(-1,0,0)$
  & \quad $\left\{3,0,3+\frac{\sqrt6}{2}\mu\right\}$
  & \quad no \\ \noalign{\vskip 1mm}
$C(0,0,0)$
  & \quad $\left\{-\frac32,\frac32,\frac32\right\}$
  & \quad no \\ \noalign{\vskip 1mm}
$D\left(\frac{\mu}{\sqrt6},\sqrt{1-\frac{\mu^2}{6}},0\right)$
  & \quad $\left\{3-\frac{\mu^2}{2},-3+\frac{\mu^2}{2},-3+\mu^2\right\}$
  & \quad no \\ \noalign{\vskip 1mm}
$E\left(\frac{\sqrt6}{2\mu},\frac{\sqrt6}{2\mu},0\right)$
  & \quad $\left\{\frac32,-\frac34\left( 1-\sqrt{\frac{24}{\mu^2}-7} \right),
  -\frac34\left( 1+\sqrt{\frac{24}{\mu^2}-7} \right) \right\}$
  & \quad no \\ \noalign{\vskip 1mm}
$F\left(\frac{\mu}{\sqrt6},\sqrt{1-\frac{\mu^2}{6}},
\frac{3(\mu^2-6)}{(2\alpha+3)\mu^2-18}\right)$
  & \quad $\left\{\frac{\alpha\mu^4}{(2\alpha+3)\mu^2-18},
  \frac{\alpha\mu^2(\mu^2-6)}{(2\alpha+3)\mu^2-18},
  \frac{\alpha\mu^2(\mu^2-6)}{(2\alpha+3)\mu^2-18}\right\}$
  & \quad no \\ \noalign{\vskip 1mm}
$G(0,y,1)$
  & \quad $\left\{0,0,\alpha(1-y^2)\right\}$
  & \quad $\mu>0$, $\alpha<0$, $y=1$ \\ \noalign{\vskip 1mm}
\multirow{2}{1cm}{\centerline{$H\left(x,\sqrt{1-x^2},1\right)$}}
& \quad
  \multirow{2}{4cm}{\centerline{$\left\{0,-2\alpha x^2,-2\alpha x^2\right\}$}}
& \quad $\mu>0$, $\alpha>0$, $|x|\leq1$ \\ 
& & \quad \mbox{or} $\mu>0$, $\alpha<0$, $x=0$ \\  \noalign{\vskip 1mm}
    \hline
\end{tabular}
\caption{\label{Table:stability CP} Stability of the critical points and
critical lines of the dynamical system~(\ref{DSn}). None of the critical 
points $A$ to $F$ is stable. For positive $\alpha$ (transfer of energy from
dark matter to dark energy), the critical line $H\left(x,\sqrt{1-x^2},1\right)$
is an attractor, while for negative $\alpha$ (transfer of energy from dark
energy to dark matter) the attractor is the point $G(0,1,1)=H(0,1,1)$, lying
at the intersection of the critical lines. }
\end{table*}

The critical points $A$, $B$, $C$, $D$, and $E$, lying on the plane $z=0$,
are those already present in the uncoupled quintessence model.
This follows from the fact that for $z=0$, Eqs.~(\ref{DSn-x}) and
(\ref{DSn-y}) decouple from Eq.~(\ref{DSn-z}), yielding the two-dimensional
dynamical system of Ref.~\cite{copeland-1998}, which describes a quintessence 
scalar field interacting only through gravity with a pressureless fluid.
The critical point $F$ and the critical lines $G$ and $H$ are new;
they arise due to the introduction of a direct interaction term between
dark energy and dark matter.

Inspection of the eigenvalues shown in Table~\ref{Table:stability CP} for
all possible values of the parameters $\alpha$ and $\mu$ belonging to the
parameter space reveals that none of the critical points $A$ to $F$ can be an
attractor, because at least one of the corresponding eigenvalues is positive.
The same applies to the critical lines $G$ for $\alpha>0$ ($y\neq1$)
and $H$ for $\alpha<0$ ($x\neq0$).
Thus, we conclude that the attractors can only be the critical lines
$G$ and $H$ for $\alpha<0$ and for $\alpha>0$, respectively.
Let us analyze in detail these two cases.

The critical points belonging to the critical line $H(x,\sqrt{1-x^2},1)$
have just one eigenvalue equal to zero (for $x\neq0$) and, therefore,
they are normally hyperbolic, meaning that the linear stability theory
suffices to assess the behavior of the trajectories in their vicinity.
Since for $\alpha>0$ the other two eigenvalues are negative,
we conclude that, for such values of the parameter $\alpha$ and $x\neq0$,
these critical points are stable, i.e.,
they attract the trajectories along the noncritical directions.
By continuity arguments, the critical point $H(0,1,1)$ is also
an attractor for $\alpha>0$.

The stability of the critical points belonging to the critical line
$G(0,y,1)$ is not so straightforward to assess.
Indeed, because at least two eigenvalues are equal to zero, the linear
theory does not suffice, and one has to resort to alternative methods to
investigate stability. 
In Appendix~\ref{Sec-appendix} we show using the center manifold theory
that, for $\alpha<0$, the trajectories, when approaching the critical line $G$,
drift in the $y$-direction, converging to the critical point
$G(0,1,1)=H(0,1,1)$, which lies at the intersection of the critical lines
$G$ and $H$.
Therefore, this point is a global attractor for $\alpha<0$.

In summary, for $\alpha>0$ the critical line $H(x,\sqrt{1-x^2},1)$
is the global attractor, while for $\alpha<0$ all trajectories
are attracted to the critical point $G(0,1,1)=H(0,1,1)$.

Taking into account that at the critical lines $G$ and $H$
the dark-energy density parameter is $\Omega_\phi=y^2$ and 
$\Omega_\phi=1$, respectively (see Table~\ref{Table:properties CP}),
from the above results on the stability of the critical points and lines,
it immediately follows that, asymptotically, all cosmological solutions
describe a Universe completely dominated by dark energy, irrespective of
the value of $\alpha$, which, recall, parameterizes the energy exchange
between the dark components of the Universe.

This result is somewhat unexpected since, at first sight, a significant
transfer of energy from the dark-energy scalar field to the dark-matter
fluid should favor the appearance of scaling solutions, i.e.,
solutions for which the ratio between the density parameters of dark matter
and dark energy is nonzero.
However, as shown above, there are no attractors corresponding to such
scaling solutions, and the asymptotic dominance of dark energy is an
unavoidable feature of the cosmological model under consideration.
Still, there is a subtlety here; we will clarify it below after presenting
a detailed description of the solutions of cosmological relevance.

Note that the critical point $G(0,1,1)=H(0,1,1)$ is deep inside the region
of the phase space in which expansion is accelerated ($x^2-y^2<-1/3$),
while the critical line $H(x,\sqrt{1-x^2},1)$ is only partially inside
this region.
This means that, for $\alpha<0$, the attractor always corresponds
to a Universe with accelerated expansion, while for $\alpha>0$,
the attractor corresponds to accelerated expansion only if $|x|<1/\sqrt3$.

Agreement with observations requires the present era of accelerated
expansion of the Universe to be preceded by a matter-dominated era,
long enough to allow for structure formation.
This sequence of cosmological eras corresponds to those trajectories
of the dynamical system~(\ref{DSn}) that pass near the critical point $C$
before proceeding to the global attractor at $G(0,1,1)=H(0,1,1)$ for 
$\alpha<0$, or $H(x,\sqrt{1-x^2},1)$ for $\alpha>0$ and $|x|<1/\sqrt{3}$.
In what follows, we will focus our attention on such trajectories
since they correspond to solutions that agree, at least qualitatively, with
cosmological observations.

Each set of values of the parameters $\alpha$ and $\mu$ corresponds to
a specific trajectory in the phase space of the dynamical system~(\ref{DSn}).
Let us recall that $\alpha$ parameterizes the direct transfer of energy between
the two dark components of the model (from dark matter to dark energy if
$\alpha>0$, in the opposite direction if $\alpha<0$) and $\mu$ parameterizes
the steepness of the scalar-field potential.

For $\alpha>0$, we can restrict our analysis to $0<\mu<\sqrt{2}$
because only for these values of the parameter $\mu$ does the final state
of evolution correspond to accelerated expansion.
In this case, the trajectories we are interested in pass close to the critical
point $C$ (close enough to guarantee a matter-dominated era of appropriate
duration), proceed to the vicinity of the critical point $D$, and then climb
vertically in the direction of the $z=1$ plane, converging asymptotically to
the point with coordinates $x=\mu/\sqrt{6}$, $y=\sqrt{1-\mu^2/6}$, and $z=1$,
belonging to the critical line $H$.
At this final state, the effective equation-of-state parameter is given by 
$w_\texttt{eff}=-1+2x^2\simeq-1+\mu^2/3$ (see Table~\ref{Table:properties CP}).

\begin{figure}[t]
 \includegraphics{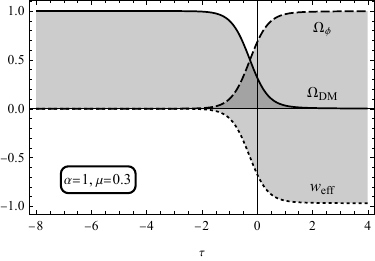}
\caption{\label{alpha positive} 
Evolution of $\Omega_\phi$, $\Omega_\texttt{DM}$, and $w_\texttt{eff}$
for the case $\alpha=1$ and $\mu=0.3$. A matter-dominated era, long enough
to allow for structure formation, is followed by an era of dark-energy
domination, during which the expansion of the Universe is accelerated.}
\end{figure}

\begin{figure}[t]
 \includegraphics{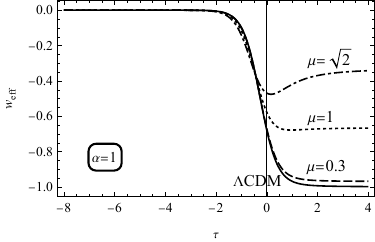}
\caption{\label{dif mu} Evolution of the effective equation-of-state parameter
$w_\texttt{eff}$ for $\alpha=1$ and different values of the parameter $\mu$.
For comparison, the curve corresponding to the $\Lambda$CDM concordance model
is also shown. For small values of $\mu$, the evolution of the coupled
quintessence cosmological model is quite similar to $\Lambda$CDM.  }
\end{figure}

The evolution of the dark-energy density parameter $\Omega_\phi$,
the dark-matter density parameter $\Omega_\texttt{DM}$, and the effective
equation-of-state parameter $w_\texttt{eff}$ is illustrated in
Fig.~\ref{alpha positive} for the case $\alpha=1$ and $\mu=0.3$.
Initial conditions $x_i=y_i=9.21\times10^{-6}$ and $z_i=10^{-7}$ were chosen
to guarantee a long enough matter-dominated era, stretching from $\tau=-8$
to $\tau=-0.4$, which corresponds to a redshift between $3000$ and $0.5$,
and also to guarantee that, at the present time $\tau=0$, 
$\Omega_\phi(0)=0.68$.
The effective equation-of-state parameter tends, asymptotically, to
$-0.97$.
In Fig.~\ref{dif mu}, the evolution of the effective
equation-of-state parameter $w_\texttt{eff}$ is shown for different values
of $\mu$.
For comparison, we also show the curve corresponding to the $\Lambda$CDM
concordance model, obtained by choosing $\alpha=0$, $\mu=0$, and $x_i=0$.

Let us now turn to the case $\alpha<0$.
For $0<\mu\leq\sqrt{3}$, the behavior of the trajectories is similar, initially,
to the case $\alpha>0$; they pass near $C$ and $D$, but then they climb to
the vicinity of the critical point $F$ (lying above $D$), and, from there,
they proceed to the final state at $G(0,1,1)=H(0,1,1)$.
For $\sqrt{3}<\mu<\sqrt{6}$, the trajectories, after passing near $C$, proceed
to $E$ (instead of $D$), then approach $F$, before heading to the attractor
at $G(0,1,1)=H(0,1,1)$.
Finally, for $\mu\geq\sqrt{6}$, the trajectories pass near $C$,
then approach $E$, from where they proceed directly to the attractor at
$G(0,1,1)=H(0,1,1)$.

\begin{figure}[t]
 \includegraphics{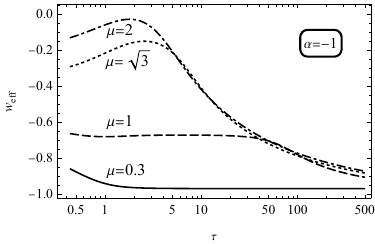}
\caption{\label{alpha negative} Evolution of the effective equation-of-state
parameter $w_\texttt{eff}$ for $\alpha=-1$ and different values of the
parameter $\mu$. For negative $\alpha$, $w_\texttt{eff}$ asymptotically
converge to $-1$ irrespective of $\mu$.}
\end{figure}

\begin{figure}[t]
 \includegraphics{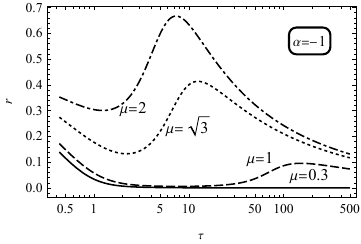}
\caption{\label{ratio} Evolution of the ratio
$r\equiv\Omega_\texttt{DM}/\Omega_\phi$ for $\alpha=-1$ and different values
of the parameter $\mu$. For negative $\alpha$, the convergence of $r$ to zero
proceeds at an exceedingly slow rate, meaning that, in the near future,
the solutions effectively behave as scaling solutions.}
\end{figure}

In all cosmological solutions with $\alpha<0$, at the final state, the
effective equation-of-state parameter is given by $w_\texttt{eff}=-y^2=-1$
(see Table~\ref{Table:properties CP}).
This corresponds to accelerated expansion irrespective of the value of the
parameter $\mu$.
This result is illustrated in Fig.~\ref{alpha negative}, where
$w_\texttt{eff}$ is seen to asymptotically converge to $-1$ for different
values of the parameter $\mu$.
Note, however, that this convergence proceeds at an exceedingly slow rate.
Meanwhile, the solution mimics a scaling solution, for which the energy
density of dark matter is a significant fraction of the total energy density,
as shown in Fig.~\ref{ratio}.
In both Figs.~\ref{alpha negative} and \ref{ratio}, initial conditions
were chosen to guarantee a long enough matter-dominated era and also
$\Omega_\phi(0)=0.68$, namely, 
$x_i=y_i=9.21\times10^{-6}$ for $\mu=0.3$,
$x_i=y_i=1.024\times10^{-5}$ for $\mu=1$, 
$x_i=y_i=1.455\times10^{-5}$ for $\mu=\sqrt{3}$,
and $x_i=y_i=2.04\times10^{-5}$ for $\mu=2$ ($z_i=10^{-7}$ in all cases).

The asymptotic behavior of the above-described cosmological solutions can be
better understood by analyzing the evolution of the interaction term $Q$,
which, in terms of the variables $x$, $y$, and $z$, reads
\begin{equation}
 Q=\frac{6\alpha H_*^3}{\kappa^2} \frac{(1-z)^2 x^2}{(1-x^2-y^2)^2 z^2},
\end{equation}
where $H_*$ is the positive constant introduced in Eq.~(\ref{xyz}).

\begin{figure}[t]
 \includegraphics{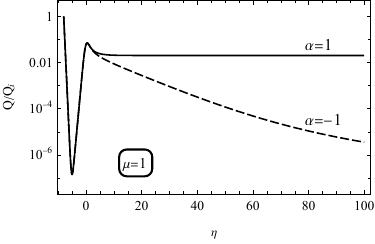}
\caption{\label{Q} Evolution of the interaction term $Q$ (divided by its
value at the beginning of the numerical integration $Q_i$). For $\alpha>0$,
a steady transfer of energy from dark matter to dark energy reinforces the
asymptotic dominance of the latter. For $\alpha<0$, an initial significant
transfer of energy from dark energy to dark matter helps the latter to be
maintained at an expressive level, a situation that mimics the behavior of a
scaling solution; however, as time passes, this energy transfer tends to zero
and the Universe is allowed to complete its transition to an era of total
dark-energy domination.}
\end{figure}

As shown in Fig.~\ref{Q}, for $\alpha>0$, $Q$ rapidly converges to a constant
positive value, which guarantees a steady transfer of energy from the
dark-matter fluid to the dark-energy scalar field and, consequently,
reinforces the dominance of the latter in the evolution of the Universe.
For $\alpha<0$, there is initially a significant transfer of energy from
dark energy to dark matter, keeping the energy density of the latter at an 
expressive level---a situation that mimics the behavior of a scaling
solution--- but, as time goes on, this transfer of energy approaches zero,
the energy density of the dark-matter fluid becomes more and more negligible,
and the Universe finally completes its transition to an era of total
dark-energy domination.
In Fig.~\ref{Q}, initial conditions were chosen, again, to guarantee a long
enough matter-dominated era and also $\Omega_\phi(0)=0.68$, namely, 
$x_i=y_i=1.024\times10^{-5}$ and $z_i=10^{-7}$.

\section{Conclusions \label{Sec-conclusions}}

In this article, we have investigated a coupled quintessence cosmological
model with an interaction term inspired by warm inflation.

With an appropriate choice of dimensionless variables, the evolution
equations of the coupled quintessence model can be written as a 
three-dimensional dynamical system.

A stability analysis of the dynamical system's critical points and lines
shows that, asymptotically, all cosmological solutions describe a Universe
completely dominated by dark energy, irrespective of the values
of the parameters $\alpha$ and $\mu$.

However, a thorough study of the phase-space trajectories 
for different $\alpha$ and $\mu$ reveals that in the case of negative
$\alpha$ (corresponding to a transfer of energy from the dark-energy
scalar field to the dark-matter fluid), for any value of $\mu$,
the approach to the final state of
dark-energy dominance is so slow that, in the near future, the cosmological
solutions effectively behave as scaling solutions.

Such behavior had already been observed previously in another coupled 
quintessence model \cite{potting-2022} with an interaction term between
dark energy and dark matter of the form
$Q\propto\rho_{\texttt{DM}} \dot{\phi} \phi^n$, where
$n$ is a positive integer. 
In that study, as in the present one, strictly speaking, there are no
scaling attractor solutions; however, for certain values of a relevant
parameter, there are solutions that, during the approach to the final state
of cosmic evolution, behave, for all practical purposes, as accelerated
scaling solutions.
It seems thus that in coupled quintessence cosmological models, scaling
solutions are more ubiquitous than one might expect from a simple stability
analysis of the critical points of the corresponding dynamical system.

The analysis of the dynamical system (\ref{DSn}) also reveals the existence
of a family of phase-space trajectories corresponding to an appropriate sequence
of cosmic eras, namely, a matter-dominated era, long enough to allow for
structure formation, followed by a present era of accelerated expansion.
Such cosmologically relevant trajectories exist for both positive and negative
values of the parameter $\alpha$ (for $\alpha>0$, accelerated
expansion at the current time requires $\mu<\sqrt2$).
Thus, irrespective of the direction of energy transfer between dark
matter and dark energy, the coupled quintessence cosmological model given
by Eqs.~(\ref{Friedmann 1})--(\ref{dotrho}) is able to replicate the observed
late-time stages of the evolution of the Universe.

Within warm inflationary models, a variety of forms have been adopted for
the dissipation coefficient, from the simplest, based on general
phenomenological considerations, to the more elaborate ones, derived from
quantum field theory.
As a first approach to the study of coupled quintessence models with an
interaction term inspired by warm inflation, we have considered $\Gamma$
in Eq.~(\ref{interaction term}) to be constant.
Dissipation coefficients depending on both the scalar field $\phi$ and the
dark-matter energy density $\rho_{\texttt{DM}}$ will be considered in
future work.

\begin{acknowledgments}
The author acknowledges support from Funda\c{c}\~ao para a Ci\^encia e a 
Tecnologia (Portugal) through the research grants doi.org/10.54499/UIDB/04434/2020
and doi.org/10.54499/UIDP/04434/2020 and thanks
A. J. Roberts for comments on central manifold theory.
\end{acknowledgments}


\appendix*
\section{Stability analysis of the critical line
$G(0,y,1)$\label{Sec-appendix}}

Consider a specific point $G(0,y_c,1)$ of the critical line $G(0,y,1)$,
where $y_c\neq1$ is a constant.

The eigenvalues of the stability matrix of the dynamical system~(\ref{DSn}),
evaluated at this point, are $\lambda_{1,2}=0$ and
$\lambda_3=\alpha(1-y_c^2)$.
Because two eigenvalues are zero, the stability properties of the critical
point cannot be determined within the linear theory; we must resort to 
alternative methods.
Here, we will use the center manifold theory to assess the
stability of the critical point $G(0,y_c,1)$
(see Refs.~\cite{sa-2021,potting-2022} for examples of this theory's
application in cosmological contexts similar to the present article).

Introducing new variables
\begin{equation}
 u=x+\frac{\sqrt6\mu y_c^2}{2\alpha(y_c^2-1)}(z-1), 
 \quad v=y-y_c, \quad w=z-1,
 \label{cv1}
\end{equation}
the dynamical system~(\ref{DSn}) is brought to the form
\begin{subequations}\label{u,v,w-eta-n=1}
\begin{align}
 u_\tau & = \alpha (1-y_c^2)u + f_1(u,v,w),
 \label{flow u}
\\
 v_\tau & = -\frac32 y_c (1-y_c^2)w + f_2(u,v,w),
\\
 w_\tau & = f_3(u,v,w),
\end{align}
\end{subequations}
where $f_i=\mathcal{O}(u^2,v^2,w^2,uv,uw,vw)$, $i=1,2,3$.

The center manifold $u=h(v,w)$, which is a solution of the partial
differential equation
\begin{align}
& \frac{\partial h}{\partial v} \left[ -\frac32 y_c (1-y_c^2)w + f_2 
	\big(h(v,w),v,w\big) \right] \nonumber
\\ 
 &\hspace{5mm} +\frac{\partial h}{\partial w} f_3\big(h(v,w),v,w\big)
  -\alpha(1-y_c^2) h(v,w)
 \nonumber
\\
 &\hspace{5mm}  -f_1\big(h(v,w),v,w\big)=0,
\end{align}
is given by
\begin{equation}
 h(v,w) = \frac{\sqrt6\mu}{2\alpha}v^2w
 +\mathcal{O}(v^4,v^3w,v^2w^2,vw^3,w^4),
 \label{CM-0}
\end{equation}
for $y_c=0$, and
\begin{align}
 h(v,w) =& \frac{\sqrt6\mu y_c}{\alpha(1-y_c^2)^2}vw
           -\frac{\sqrt6\mu y_c^2[3+\alpha(1-y_c^2)]}
                 {2\alpha^2(1-y_c^2)^2} w^2  \nonumber
\\
                 &+\mathcal{O}(v^3,v^2w,vw^2,w^3),
 \label{CM}
\end{align}
for $0<y_c<1$. 

The flow on the center manifold is determined by the differential equations
\begin{equation}
v_\tau =  -\frac32 vw \quad \mbox{and} \quad w_\tau =  \frac32 w^2,
\end{equation}
for $y_c=0$, and
\begin{equation}
v_\tau = -\frac32 y_c (1-y_c^2)w \quad \mbox{and} \quad
w_\tau = \frac32 (1+y_c^2) w^2,
\end{equation}
for $0<y_c<1$. In the $u$-direction the flow is given by Eq.~(\ref{flow u}).

Taking into account that in the neighborhood of the critical point
$w<0$ (also $v>0$ for the case $y_c=0$),
we conclude that both $v_\tau$ and $w_\tau$ are positive, implying that the
trajectories approach the critical point along the $w$-direction and move away
from it along the $v$-direction. 
Along the $u$-direction, the trajectories approach the critical point if
$\alpha<0$ and move away from it if $\alpha>0$.

In terms of the original variables $x$, $y$, and $z$, the above results mean
that, for $\alpha<0$, the trajectories, when approaching the critical line
$G(0,y,1)$, drift in the $y$-direction, asymptotically converging to the
critical point $G(0,1,1)=H(0,1,1)$, which lies in the intersection of
the critical lines $G$ and $H$. 
Hence, this point is a global attractor for $\alpha<0$.

\end{document}